\documentclass[aoas,MSNbibl,nameyear,dvips]{arximspdf}

\doi{10.1214/13-AOAS655} 
\volume{7}
\issue{3}
\pubyear{2013}
\firstpage{1796}
\lastpage{1813}

\makeatletter

\setattribute{copyright}{owner}{\textup{In the Public Domain}}

\newtheorem{theorem}{Theorem}

\newcommand{\CP}{\mathit{CP}}
\newcommand{\TC}{\mathit{TC}}

\makeatother

\begin{document}
\begin{frontmatter}

\title{Likelihood reweighting methods to reduce potential bias in
noninferiority trials which rely on historical data to make
inference\thanksref{T1}}
\runtitle{Likelihood reweighting methods}

\thankstext{T1}{This article reflects the views of the authors and
should not be construed to represent the FDAs views or policies.}

\begin{aug}
\author[A]{\fnms{Lei} \snm{Nie}\corref{}\ead[label=e1]{lei.nie@fda.hhs.gov}},
\author[B]{\fnms{Zhiwei} \snm{Zhang}\ead[label=e2]{zhiwei.zhang@fda.hhs.gov}},
\author[A]{\fnms{Daniel} \snm{Rubin}}
\and
\author[B]{\fnms{Jianxiong} \snm{Chu}}
\runauthor{Nie, Zhang, Rubin and Chu}
\affiliation{The U.S. Food and Drug Administration}
\address[A]{L. Nie\\
D. Rubin\\
Division of Biometrics IV\\
Office of Biostatistics/CDER/FDA\\
10903 New Hampshire Avenue\\
Silver Spring, Maryland 20993\\
USA} 
\address[B]{Z. Zhang\\
J. Chu\\
Division of Biostatistics\\
Office of Surveillance\\
\quad and Biometrics/CDRH/FDA\\
10903 New Hampshire Avenue\\
Silver Spring, Maryland 20993\\
USA\\
\printead{e2}}
\end{aug}

\received{\smonth{2} \syear{2013}}
\revised{\smonth{4} \syear{2013}}

%
\begin{abstract}
It is generally believed that bias is minimized in well-controlled
randomized clinical trials. However, bias can arise in active
controlled noninferiority trials because the inference relies on a
previously estimated effect size obtained from a historical trial that
may have been conducted for a different population. By implementing a
likelihood reweighting method through propensity scoring, a study
designed to estimate a treatment effect in one trial population can be
used to estimate the treatment effect size in a different target
population. We illustrate this method in active controlled
noninferiority trials, although it can also be used in other types of
studies, such as historically controlled trials, meta-analyses, and
comparative effectiveness analyses.
\end{abstract}

%
\begin{keyword}
\kwd{Bias}
\kwd{generalized linear model}
\kwd{inverse probability weighting}
\kwd{noninferiority}
\kwd{propensity score}
\end{keyword}

\end{frontmatter}

\section{Introduction}
\label{sec1}

The code of federal regulations (CRF) Chapter 21, Section~314.126, states
that ``The purpose of conducting clinical investigations of a drug is to
distinguish the effect of a drug from other influences$\ldots$'' and the
purpose is achieved through ``adequate and well-controlled clinical
investigation.'' \mbox{According} to the CRF, an adequate and well-controlled
trial has a number of characteristics, including: (1) ``The method of
assigning patients to treatment and control groups minimizes bias and is
intended to assure comparability of the groups with respect to pertinent
variables such as age, sex, severity of disease$\ldots$'' and (2) ``Adequate
measures are taken to minimize bias on the part of the subjects,
observers$,\ldots\,$.'' Characteristic (1) and part of (2) aim to minimize
bias through balancing the population between the two treatment arms.

By conducting well-controlled clinical trials, we generally anticipate that
systematic bias is minimized in superiority trials. However, this belief
may be more tenuous in noninferiority trials. Note that noninferiority
trials are the major vehicle to evaluate new treatments in many disease
areas, after the pioneering consideration of ethical issues in
Placebo-controlled trials by \citet{RotMic94}.

Consider a Palivizumab-controlled noninferiority trial of Motavizumab
for prophylaxis of serious respiratory syncytial virus (RSV) disease in
high risk children [\citet{Caretal10}]. This trial will be called
MOTA throughout the paper. The goal of the trial was to evaluate
whether Motavizumab was noninferior to Palivizumab in the rate of
hospitalization attributed to RSV. Let $\hat{\mu}_{\TC}$ be the
estimated log-odds ratio of Palivizumab vs. Motavizumab, and let
$\hat{\mu}_{\CP}$ be the estimated log-odds ratio of Placebo vs.
Palivizumab. Because the log-odds ratio of Placebo vs. Motavizumab
cannot be estimated directly (the noninferiority trial does not have a
placebo arm), $\hat{\mu}_{\TC}+\hat{\mu}_{\CP}$ is often used as an
indirect estimate, with a standard error of $\sqrt{\sigma_{\TC}^{2} +
\sigma_{\CP}^{2}}$. We may consider the noninferiority of Motavizumab
to Palivizumab to be met at level $\alpha$ if \mbox{$\hat{\mu} =
(\hat{\mu}_{\TC} + \hat{\mu}_{\CP})/\sqrt{\sigma_{\TC}^{2} +
\sigma_{\CP}^{2}} > Z_{\alpha}$}, where $Z_{\alpha}$ is the ($100 -
\alpha$)th percentile of a standard normal distribution. In this
example, $\hat{\mu}_{\CP}$ and its standard error $\sigma_{\CP}$ were
obtained from an earlier Placebo-controlled trial of Palivizumab,
\citet{autokey12}, in which $\hat{\mu}_{\CP}=0.86$ (corresponding
to odds ratio of~2.4) with a standard error of 0.21. This trial will be
called IMPACT throughout the paper. Now, keeping in mind the fact that
the statistics $\hat{\mu}$ synthesizes $\hat{\mu}_{\CP}$ and
$\hat{\mu}_{\TC}$, with the former estimated from the IMPACT population
and the latter estimated from the MOTA population, we illustrate the
following issues. First, both MOTA and IMPACT enrolled subjects
exclusively from two disjoint subgroups: (1) children $\leq$24 months
with a clinical diagnosis of Bronchopulmonary dysplasia (BPD); and (2)
children with $\leq$35 weeks gestation and $\leq$6 months, who did
not have BPD. The proportion of subjects with BPD was 51\% in IMPACT
and only 22\% in MOTA. Second, treatment heterogeneity of Palivizumab
was observed in these two subgroups in IMPACT. For subjects enrolled
with BPD, the odds ratio was 4.88 with a 95\% C.I. of $(2.17, 10.96)$,
and for subjects enrolled without BPD, the odds ratio was 1.72 a 95\%
C.I. of $(1.06, 2.79)$ (see Section \ref{sec4}). The Wald Chi-square
test of treatment by BPD interaction through a logistic regression was
significant with a $p$-value of 0.03. Because of the population
difference and treatment heterogeneity, an appropriate odds ratio
$\hat{\mu}_{\CP}$ used in $\hat{\mu}$ should reflect the population of
MOTA, while the value of 0.86 instead reflects the population of
IMPACT. Using data provided in Section \ref{sec4}, we obtain the
adjusted incidence rate of Placebo in the MOTA population of $34/266
\times22\mbox{\%} + 19/234 \times78\%=9.2\%$, and the adjusted
incidence rate of Palivizumab of $39/496 \times22\% + 9/506 \times
78\%=3.1\%$. Therefore, the adjusted log-odds ratio is 1.14 and the
adjusted odds ratio is 3.1. Consequently, the adjusted log-odds ratio
of Placebo vs. Palivizumab in the MOTA population should be better
quantified as 1.14 rather than 0.86, the unadjusted log-odds ratio
$\hat{\mu}_{\CP}$. The difference between 1.14 and 0.86 is a bias
associated with this inference.

In the previous example, it was easy to adjust for the population
difference, which only involved heterogeneity in BPD status. In some other
examples, the situation could be more complicated. For example, in the
development of Elvitegrevir [\citet{Moletal12}], the trial population
was different from the historical trial population in several
characteristics, for which treatment heterogeneity has been reported
[\citet{Cooetal08}].

These examples show that analysis of a noninferiority trial may rely on a
combination of information from the trial itself and one or more historical
trials. The main issue is that the populations of the noninferiority and
historical trials may be different. If treatment heterogeneity is present,
an inference that does not adjust for the population difference can be
biased.

Covariate adjustment approaches [\citet{Zha09} and \citet
{NieSoo10}] have
been proposed to address the problem. Both approaches involve a regression
model relating the clinical outcome to treatment and relevant covariates.
They cannot be directly applied to obtain the marginal (crude) odds ratio,
which is the prespecified primary endpoint in the aforementioned
examples.

This paper proposes a calibration method through likelihood reweighting so
that a study designed to estimate a marginal treatment effect size for one
trial population (e.g., IMPACT) may be used to calibrate the effect
size in
a different but closely related study population (e.g., MOTA). We prove
that the maximum likelihood estimator for this reweighted likelihood is a
consistent estimator of the treatment effect size in the targeted
population. In addition, we also propose a nonparametric approach based on
the calibration method.

The proposed calibration approach using the likelihood reweighting method
is (asymptotically) equivalent to the covariate adjustment approach in some
cases such as linear regression, however, they are different in other
cases. The choice between the two approaches can be subtle and subjective.
An important consideration is to make sure that $\hat{\mu}_{\TC}$ and
$\hat{\mu}_{\CP}$ in the statistics $\hat{\mu}_{ks} = (\hat{\mu}_{\TC} +
\hat{\mu}_{\CP})/\sqrt{\sigma_{\TC}^{2} + \sigma_{\CP}^{2}}$ have similar
interpretations. Specifically, if $\hat{\mu}_{\TC}$ is a marginal (i.e.,
overall) treatment effect as in the previous two examples and in most
randomized clinical trials, then $\hat{\mu}_{\CP}$ should probably be
calibrated using the method presented in this paper so as to maintain the
marginal interpretation. In Section~\ref{sec3.3} we also make some
observations on
the likelihood reweighing method as an alternative to the covariate
adjustment approach used in randomized clinical trials, along with the
differences noted in the literature.

Although this paper mainly targets noninferiority trials, the results are
also applicable to historically controlled trials, which have similar
issues [\citet{FriFurDem98}]. A comparison of the likelihood reweighting
method to related methods in historically controlled trials, for example,
\citet{Zha07}, \citet{Sigetal10} and \citet{Sigetal11}, is
provided in the supplement [\citet{NieZhaRub}]. This paper focuses on
calibrating the treatment effect size from one population to another
population which is different but overlapping. It is related to but
different from studies generalizing results from a subpopulation to a
strictly larger population (whole population); see \citet{ColStu10},
\citet{Greetal08}, \citet{WeiHayPon09} and \citet{Fra09},
among others. These references are restricted to the clinical trial
literature, although other areas, such as observational studies, involve
similar problems.

\section{Motivation, assumptions, and notation}
\label{sec2}

\subsection{Motivation} The idea behind our method is simple. Recall the
\hyperref[sec1]{Intro-} \hyperref[sec1]{duction} where we obtained the expected incidence rate of Placebo in
the MOTA population as
\[
12.8\% \times22 \% + 8.1\% \times78\%
= \frac{\sum_{i = 1}^{500}
y_{i}p_{x_{i},\mathrm{MOTA}}/p_{x_{i},\mathrm{IMPACT}}} {500},
\]
where $p_{x_{i},\mathrm{MOTA}}$ and $p_{x_{i},\mathrm{IMPACT}}$ are the
percentage of Placebo
subjects with characteristics $x_{i}$ in MOTA and IMPACT trials,
respectively. When $x_{i}=1$ (a~diagnostic of BPD), the
$p_{x_{i},\mathrm{MOTA}}$ and
$p_{x_{i},\mathrm{IMPACT}}$ are 22\% and 53\%, respectively; when
$x_{i}=0$, they
are 78\% and 47\%. By defining $p_{x_{i},\mathrm{MOTA}}/\break p_{x_{i},\mathrm
{IMPACT}}$ as $r_{i}$,
the expected incidence rate of Placebo in the MOTA population is simply the
mean of reweighted response from all subjects and the weight reflects the
change of population difference with respect to BPD status.

In many other situations, the parameters cannot be directly calibrated as
shown in this example. They can, however, be estimated using a likelihood
approach to be described shortly.

Robins and colleagues gave an intuitive explanation of how the inverse
probability weighting approach reduces bias in the context of estimating
marginal structural models (MSMs) in epidemiology [\citet{RobHerBru00}].
Heuristically, weighting each subject by the inverse of the propensity
score for the treatment actually received creates a confounding-free
pseudo-population, where treatment assignment is independent of the
potential outcomes. Typically, the inverse probability weighting approach
is used to estimate marginal means of potential outcomes in an estimating
equation framework. However, the insights of the work by Robins and
colleagues certainly extend to likelihood-based inference and allow us to
calibrate the treatment effect. Specifically, upon appropriately
reweighting the likelihood function contributed by each subject, a
calibrated treatment effect can be obtained. Before illustrating the
reweighted likelihood approach, we introduce some notation and
assumptions.

\subsection{Assumptions and notations} Consider a trial conducted in a
population~$P$ (e.g., the IMPACT population) to compare treatment 1 (e.g.,
Palivizumab) to treatment 2 (e.g., Placebo). We assume that a random sample
from $P$ is randomly assigned into these two treatment groups. The
objective of the trial is to quantify the treatment effect size of
treatment 1 relative to treatment 2 in population $P$. The objective of
this paper is to calibrate the effect size of treatment 1 relative to
treatment 2 from the original population to a different but closely related
population~$P^*$ (e.g., MOTA population).

We assume that the populations $P$ and $P^*$ are different. In our first
example, $P$ refers to a population comprised of subjects with BPD (51\%)
and without BPD (49\%) and $P^*$ refers to a population with a different
composition (22\% with BPD, 78\% without BPD).

We also assume that the populations $P$ and $P^*$ are closely related and
that the differences between $P$ and $P^*$ are entirely captured by the
value of a predictive covariate (vector) $X$ representing subjects'
baseline disease characteristics. In addition, we assume that all subjects
with covariate value $X=x$ are expected to have the same treatment effect,
regardless of their origin (population $P$ or population~$P^*$). That is,
subjects with the same covariate value $X$ are exchangeable in $P$ and
$P^*$. In our first example, this means subjects with the same BPD
diagnostic status, whether in the IMPACT population or the MOTA population,
are exchangeable in terms of response to treatments.

The difference and close relationship between population $P$ and $P^*$ is
further illustrated in mathematical form below after we clearly state the
objective of the paper. Let $Y$ be the response variable. We write
$\mu_{t}(X) = E ( Y|X,T = t )$ for the conditional mean response
of subjects with covariate $X$ who were assigned into treatment $T=t$, and
$\mu_{tP} = \nu [ E_{X \in P} \{ \mu_{t}(X) \} ]$ for
the transformed marginal mean response with respect to population $P$. When
$\nu(\cdot)$ is the identity function, $\mu_{tP}$ is the marginal
mean; when $Y$ is a binary variable and $\nu(\cdot)$ is the logit
function, $\mu_{tP}$ is the log odds in the population $P$. $\mu_{tP}$
may be
used to quantify the response of treatment $T=t$ from a historical trial,
although this is not the focus of this paper but a by-product. Instead this
paper focuses on noninferiority trials, in which we are interested in the
treatment effect of treatment 1 vs. treatment 2. We thus consider $\mu_{P}
= \pi [ E_{X \in P} \{ \mu_{1}(X) \},E_{X \in P} \{
\mu_{2}(X) \} ]$ as a metric to measure treatment effect of
treatment 1 vs. treatment 2.

In the historical trial (e.g., IMPACT), $\mu_{tP}$ or $\mu_{P}$ is
estimated. However, the objective in this paper is to estimate $\mu
_{tP^*} =
\nu [ E_{X \in P^*} \{ \mu_{t}(X) \} ]$ or $\mu_{P^*} =
\pi [ E_{X \in P^*} \{ \mu_{1}(X) \},E_{X \in P^*} \{
\mu_{2}(X) \} ]$ through calibration, without conducting a
different trial in population $P^*$ (MOTA population).

Let $F(x)$ and $F^*(x)$ denote the cumulative distribution functions of $X$
in $P$ and~$P^*$, respectively, and let $f(x)$ and $f^*(x)$ be the
corresponding probability density functions. We first assume that
$f^*(x)/f(x) \ne1$ for some $X=x$, which illustrates the differences
between populations $P$ and $P^*$. We also assume that $\infty> r(x) =
f^*(x)/f(x)$ is well defined. Because the populations are fully
described by
$X$, this assumption means that any subject included in $P^*$ should have
representatives with the same measurements in population $P$. This
highlights the close relationship between $P$ and $P^*$. When a value
of $x$
does not present in $P^*$, then $r(x)=0$. In this case, we shall not
use the
subjects in the historical trials with value $x$.

\section{Calibration of treatment effect size through likelihood
reweighting}
\label{sec3}

In our first example only the BPD status is considered and the weight is
easy to define. However, in our second example many predictive covariates
may need to be considered. In the latter case, the definition of the weight
is straightforward using the concept of the propensity score
[\citet{RosRub83}].

\subsection{Parametric approach}\label{sec3.1}

Assume two random samples of size $n_{1}, n_{2}$ from population $P$ are
assigned into treatment 1 and treatment 2, respectively. We assume $y_{it}$,
the ith subject's response from treatment group $t$, follows a generalized
linear model (GLM) with canonical link,
%
%
\begin{equation}\label{eq1}
l_{t}(y,\theta_{tx}) = \exp\biggl\{ \frac{y\theta_{tx} - b (
\theta_{tx} )}{a_{tx} ( \varphi_{tx} )} +
c_{tx} ( y,\varphi_{tx} ) \biggr\}.
\end{equation}

Let $g(\cdot)$ be the canonical link function; then $\mu_{t}(X) = g^{ -
1} (
\theta_{t} )$. We assume that $g(\cdot)$ is a monotone function with
continuous second derivative functions. One possible metric to measure the
treatment effect is $E_{X} ( \theta_{tx} )$ and another possible
metric is $\mu_{tP} = g [ E_{X \in P} \{ \mu_{t}(X) \}
]$. In the binomial-logistic regression case, we implicitly assume
that the log odds is additive for the metric $E_{X} ( \theta_{tx}
)$ and assume the proportion is additive for the metric $\mu_{tP} =
g [ E_{X \in P} \{ \mu_{t}(X) \} ]$. The former metric
was used in the covariate adjustment approach of \citet{NieSoo10}. The
latter metric shall be used in the likelihood reweighting method, as
introduced in this paper. These two metrics are related but usually are not
identical in nonlinear models.

To estimate $\mu_{tP}$, we construct the likelihood function
%
%
\begin{equation}
\label{eq2} \prod_{i = 1}^{n_{t}}
l_{t}(y_{it},\alpha_{t}).
\end{equation}

The maximum likelihood estimate (MLE) $\hat{\alpha}_{t}$ of $\alpha
_{t}$ is
a consistent estimate of the treatment effect size $\mu_{tP} = g [ E_{X
\in P} \{ \mu_{t}(X) \} ]$. The proof for this is standard
and similar to that of Theorem \ref{theo1} below, and is therefore omitted. However,
in this paper, our goal is to provide a consistent estimate of $\mu
_{tP^*} =
g [ E_{X \in P^*} \{ \mu_{t}(X) \} ]$. Our strategy is
to ``tilt'' the population $P$ so that it matches the population $P^*$ and
our matching tool is the propensity score.

In the likelihood (\ref{eq2}) we reweight the contribution of the likelihood
function from the ith subject from the historical trial with the weight
$r(x)$, and form a new likelihood function (\ref{equZV})
\renewcommand{\theequation}{2*}
\begin{equation}\label{equZV}
\prod_{i = 1}^{n_{t}} \bigl\{
l_{t}(y_{it},\alpha_{t}) \bigr\}^{r(x_{i})}.
\end{equation}

\begin{theorem}\label{theo1}
$\hat{\alpha}_{t}^{*}$, the MLE which maximize (\ref{equZV}), is a
consistent estimate of $\mu_{tP^*} = g [ E_{X \in P^*} \{ \mu_{t}(X)
\} ]$. In addition, $\hat{\alpha}_{t}^{*}\sim N (
\mu_{iP^*},A^{ - 1} ( \mu_{iP^*} )B ( \mu_{iP^*} )\*A^{ -
1} ( \mu_{iP^*} ) )$, where
\begin{eqnarray*}
A ( \alpha_{t} ) &=& E \biggl\{ r(x)\frac{d^{2}\log
l_{t}(y_{it},\alpha_{t})}{d\alpha_{t}^{2}} \biggr\},
\\
B ( \alpha_{t} ) &=& E \biggl\{ r^{2}(x)\frac{d\log
l_{t}(y_{it},\alpha_{t})}{d\alpha_{t}}
\frac{d\log_{t}(y_{it},\alpha_{t})}{d\alpha_{t}} \biggr\}.
\end{eqnarray*}
\end{theorem}

The proof of Theorem \ref{theo1} is given in the \hyperref[app]{Appendix}.
Theorem \ref{theo1} indicates that the
calibrated treatment effect converges to the treatment effect that
would be
presented in population $P^*$. In other words, the likelihood function
(\ref{equZV}) is reweighted in such a way that the units can be treated as
randomly sampled from a target population, not the population of the study.
Note that, if the two trials have the same population, then $r(x ) =
1$, so
that likelihood function (\ref{equZV}) reduces to (\ref{eq2}).

Briefly, we note that the parametric approach easily extends to include
some key covariates, including a treatment indicator as typically used in
noninferiority trials, in the likelihood (\ref{equZV}) as follows:
\[
\prod_{i = 1}^{n} \bigl\{ l(y_{i},
\alpha+ \beta z) \bigr\}^{r(x_{i})},
\]
where $l(y_{i},\alpha+ \beta z)$ is the likelihood function
contributed by
the ith subject and $z$ is a vector of treatment and/or covariate of
interest. The MLE converges to the parameters in the target population
$P^*$.

\subsection{Nonparametric approach}\label{sec3.2}

Section \ref{sec3.1} is based on the model assumption~(\ref{eq1}). In this
subsection we
take a nonparametric approach similar to the reweighting method of
\citet{Zha07} [see also \citet{Sigetal10} and \citet{Sigetal11}]
for a historical control problem, and estimate $E_{X \in P^*} \{
\mu_{t}(X) \}$ by $\hat{\delta}_{t} = \sum_{i = 1}^{n_{t}}
y_{it}r ( x_{i} ) /\sum_{i = 1}^{n_{t}} r ( x_{i} )$.
When $n_{t} \to\infty$,
\[
\frac{\sum_{i = 1}^{n_{t}} y_{it}r ( x_{i} )} {n_{t}} \to E_{X
\in P} \biggl\{ \mu_{t}(X)
\frac{f^* ( X )}{f ( X )} \biggr\} = E_{X \in P^*} \bigl\{ \mu_{t}(X)
\bigr
\} = \mu_{iP^*}.
\]

Here we used the fact that $r(x) = f^*(x)/f(x)$, shown in the proof of
Theorem~\ref{theo1} in the \hyperref[app]{Appendix}. Similarly, $\sum_{i =
1}^{n_{t}} r ( x_{i}
) /n_{t} \to E_{X} \{ f^* ( x )/f ( x )
\} = 1$. Therefore, $\sum_{i = 1}^{n_{t}} y_{it}r ( x_{i} )
/n_{t} \to E_{X \in P^*} \{ \mu_{t}(X) \}$ and, thus,
\[
\mu_{tP} =
\nu \bigl[ E_{X \in P} \bigl\{ \mu_{t}(X) \bigr\} \bigr]
\]
can be estimated by $\nu ( \hat{\delta}_{t} ) = v \{ \sum_{i =
1}^{n_{t}} y_{it}r ( x_{i} ) /\sum_{i = 1}^{n_{t}} r ( x_{i} ) \}$. The
variance of the estimator and therefore the confidence interval for the
desired parameter can be obtained, for example, through the bootstrap
method proposed in \citet{Efr81}.

\subsection{Likelihood reweighting method vs. the previous covariate
adjustment approach}\label{sec3.3}

Aside from the differences between two approaches previously mentioned in
the \hyperref[sec1]{Introduction}, we have the following observations on the likelihood
reweighing method as an alternative to the covariate adjustment approach
used in randomized clinical trails, along with the differences noted in the
literature.

In the covariate adjustment approach, only the covariates interacting with
treatment are considered influential and relevant to the adjustment.
However, there are other types of ``influential'' covariates. One type of
``influential'' covariates relates to noncollapsibility, as illustrated in
Table 1 from \citet{GrePeaRob99}. Assuming that Table 1 represents the
population $P$ and that $X$ and $Z$ represent the treatments and status of
a disease, 50\% of enrolled subjects have a certain disease and the other
50\% of them do not have it. The event rates are 40\% and 20\% for
treatment $1$ $(X=1)$ and $2$ $(X=0)$, respectively, in subjects with the
disease and are 80\% and 60\% in subjects without the disease. The
treatment 1 vs. treatment 2 odds ratio is 2.67 whether subjects have the
disease or not, hence no treatment heterogeneity (i.e., no treatment by
covariate interaction when measured in odds ratio). Consider a population
$P^*$, in which 86\% of enrolled subjects have the disease and the other
14\% of them do not have it. The overall odds ratio in population
$P^*$ is thus 2.44. While the covariate adjustment approach would find that
2.67 is the odds ratio of treatment 1 vs. treatment 2 in $P$ and $P^*$, the
likelihood reweighting method would find that 2.25 and 2.44 are the odds
ratio in $P$ and $P^*$, respectively. In other words, the covariate
adjustment approach estimates the conditional odds ratio and the likelihood
reweighting method estimates the marginal odds ratio.

In this example, because the conditional odds ratio is not the same as the
marginal odds ratio, the issue of noncollapsibility arises, leading to some
questions on using the odds ratio as the measure of treatment effect. The
likelihood reweighting method provides a flexible way to circumvent this
difficulty. With the percentage difference as an alternative metric
(measure of treatment effect), we could work with $\tau_{P^*} = \mu
_{1P^*} -
\mu_{2P^*}$ and $\mu_{tP^*} = \nu [ E_{X \in P^*} \{ \mu_{t}(X)
\} ]$, where $\nu (\cdot)$ is the identity function.

The covariate adjustment approach relies on the ability of the data to
detect treatment effect heterogeneity, that is, the treatment by covariate
interaction. However, in some situations, trials may not be large
enough to
detect moderate interactions because they are not designed for that
purpose. Even if some trials are large, the rarity of events could hamper
the ability to detect all heterogeneity. Therefore, it is likely that some
important interactions are not going to be detected, leading to partial
adjustment with a residual bias. In these scenarios, the likelihood
reweighting method can be a good alternative, as it estimates the marginal
effect by maximizing the ``unadjusted'' likelihood (\ref{equZV}). Although
modeling was also used to estimate propensity scores, the dependent
variable is the trial indicator, not the outcome of interest.

The covariate adjustment approach may face co-linearity problems when there
are many covariates. When that happens, the model-based adjustment requires
a difficult decision on how to omit some covariates and select a good
model. When this problem occurs, the likelihood reweighting method can
be a
good alternative, as it is less susceptible to this issue: while
co-linearity may also cause problems with parameter estimation in the
propensity score models, it does not adversely affect prediction of the
propensity score itself.

The covariate adjustment approach utilizes the same outcome data for both
model selection and formal inference based on the chosen model, and the
results could be too optimistic. In case this becomes a concern, the
likelihood reweighting method can be considered as an alternative, as
propensity score modeling and model selection do not involve outcome data.
The techniques presented here are expected to be useful in uncontrolled
observational studies as well. Although uncontrolled observational studies
inherit many more issues than these controlled trials, the difference or
change in the patient populations associated to different comparing groups
remains one of the key issues.

As pointed out by the reviewers, a possible weakness of both approaches is
that they require subject-level data from the historical control trial.
Readers are referred to \citet{NieSoo10} for some discussions. One
possible solution could be defining the weight based on summary statistics,
an idea as illustrated in \citet{Sigetal10} and \citet{Sigetal11}.

\section{Applications}\label{sec4}

In noninferiority trials, we aim to calibrate the effect size of the active
control (e.g., Palivizumab) relative to Placebo from the historical trial
population $P$ (e.g., IMPACT) to the noninferiority trial population $P^*$
(e.g., MOTA). Using Bayes' rule, we obtain $r(x) \propto\Pr(P^*|X =
x)/\Pr
(P|X = x)$. As the population $P^*$ is associated with the experimental
treatment (e.g., Motavizumab) and the population $P$ with Placebo, $r(x)
\propto\Pr(T = \mathrm{MOTA}|X = x)/\Pr(T = \mathrm{Placebo}|X = x)$, that
is, $r(x)$ is
proportional to the odds through the propensity score.

\subsection{Development of Motavizumab, a second generation of
Palivizumab}\label{sec4.1}

Pa\-livizumab is a humanized monoclonal antibody, approved and marketed for
passive immunoprophylaxis of respiratory syncytial virus (RSV) in infants
at risk for serious RSV disease. It was studied in the IMPACT trial, a
phase III randomized, double-blind, Placebo-controlled clinical trial that
was conducted to evaluate the ability of prophylaxis with Palivizumab to
reduce respiratory syncytial virus infection in high-risk infants. A total
of 1502 children with prematurity or bronchopulmonary dysplasia (BPD), also
called chronic lung disease (CLD) in infancy, were randomized to receive
either Palivizumab or Placebo intramuscularly. The primary endpoint was RSV
related hospitalization within 150 days since administration of the first
dose of treatment. For more information of this trial please refer to the
\citet{autokey12}. This trial enrolled subjects exclusively
from two disjoint subgroups: (1) children 24 months old or younger with a
clinical diagnosis of BPD requiring ongoing medical treatment; and (2)
children with 35 weeks gestation or less and 6 months old or younger, who
did not have a clinical diagnosis of BPD.

Among subjects enrolled with a diagnosis of BPD, the incidence rate of
RSV-related hospitalization was 12.8\% ($34/266$) in the Placebo arm and
7.9\% ($39/496$) in the Palivizumab arm. Among subjects enrolled
without a
diagnosis of BPD, the incidence rate of RSV-related hospitalization was
8.1\% ($19/234$) in the Placebo arm and 1.8\% ($9/506$) in the
Palivizumab arm.
Among the 500 subjects who received Placebo, 53 (10.6\%) had an RSV-related
hospitalization; among the 1002 subjects who received Palivizumab, 48
(4.8\%) had an RSV-related hospitalization (see Table \ref{tab1} for details).
It is
clear that the treatment effect of Palivizumab vs. Placebo was better in
subjects enrolled without a diagnosis of BPD than in subjects enrolled with
a diagnosis of BPD. The overall treatment effect size of Palivizumab, as
measured in the odds ratio of the Placebo vs. Palivizumab, was 2.4 with a
95\% C.I. of $(1.6, 3.5)$.

%
\begin{table}
\tablewidth=172pt
\caption{Subject distribution: numbers of subjects (numbers of
RSV-hospitalizations)}
\label{tab1}
\begin{tabular*}{\tablewidth}{@{\extracolsep{\fill}}lcc@{}}
\hline
\textbf{IMPACT trial} & \textbf{BPD} & \textbf{Non-BPD}\\
\hline
Placebo & 266 (34) & \hphantom{0}234 (19)\\
Palivizumab & 496 (39) & \hphantom{0}506 (9)\hphantom{0}\\
MOTA trial & & \\
Palivizumab & 723 (28) & 2607 (34)\\
Motavizumab & 722 (22) & 2583 (24)\\
\hline
\end{tabular*}
\end{table}

To evaluate Motavizumab, a second generation version of
Palivizumab,
a~phase~3, randomized, double-blind, Palivizumab-controlled,
multi-center, multinational noninferiority trial (MOTA) was conducted
to assess whether Motavizumab was noninferior to Palivizumab. More
precisely, the question was whether Motavizumab is at least not too
much worse than Palivizumab in the sense that the difference of
Motavizumab vs. Palivizumab is greater than the difference of Placebo
vs. Palivizumab. With the risk difference metric this means that the
rate difference of RSV hospitalization between Motavizumab and
Palivizumab is smaller than the rate difference of RSV hospitalization
between Placebo and Palivizumab. In the metric of odds ratio this means
that the odds ratio between Motavizumab and Palivizumab is smaller than
the odds ratio between Placebo and Palivizumab. In order to evaluate
noninferiority, one possible test statistic is
\[
\hat{\mu}_{ks} = \frac{\hat{\mu}_{\TC} +
\hat{\mu}_{\CP}}{\sqrt{\sigma_{\TC}^{2} + \sigma_{\CP}^{2}}},
\]
where $\hat{\mu}_{\TC}$ is the overall log-odds ratio of Palivizumab vs.
Motavizumab and $\hat{\mu}_{\CP}$ is the overall log-odds ratio of
Palivizumab vs. Placebo.

\subsection{Calibrated effect size of Palivizumab vs. Placebo in the new
MOTA study population}\label{sec4.2}

Assume that $Y_{ixt}$, the incidence of RSV hospitalization of the ith
subject, follows a logistic regression model, $y_{ixt}\sim
\operatorname{Binomail}(1,p_{xt});\break\operatorname{logit}(p_{xt}) = \theta_{xt}$ with
$x=0, 1$, representing subjects enrolled without and with a diagnosis of
BPD and $t=0,1$ representing Placebo and Palivizumab.

Whether to make inference on $p_{xt}$ (the incidence rate) or $\theta_{xt}$
(the log odds of an event) is generally subjective. Both are used
extensively in noninferiority trials. In IMPACT and MOTA, the log-odds
ratio was the primary metric, but the risk difference is the primary metric
in current HIV trials. Therefore, we shall illustrate both metrics in the
Motavizumab example.

Let us first consider quantifying the treatment effect using the risk
difference. Let $p_{n}$ denote the proportion of subgroup with $x=1$ in the
target population (e.g., MOTA population) and $p_{h}$ denote the proportion
of subgroup with $x=1$ in the historical population (e.g., IMPACT
population). It is easy to show that the MLEs of likelihood in (\ref{eq2}) and
(\ref{equZV}) are
\[
\hat{\alpha}_{t} = \frac{n_{1t}\bar{y}_{\cdot1t} + n_{0t}\bar{y}_{\cdot
0t}}{n_{1t}
+ n_{0t}};\qquad \hat{\alpha}_{t}^{*}
= \frac{n_{1t}({p_{n}}/{p_{h}})\bar{y}_{\cdot1t} + n_{1t}(({1 -
p_{n}})/({1 -
p_{h}}))\bar{y}_{\cdot0t}}{n_{1t}({p_{n}}/{p_{h}}) + n_{1t}({1 -
p_{n}})/({1 -
p_{h}})}.
\]

In our example, these are
\begin{eqnarray*}
\hat{\alpha}_{0} &=& \frac{266 \times12.8\% + 234 \times8.1\%} {500}
= 10.6\%;\\
\hat{\alpha}_{0}^{*} &=& \biggl({266\frac{0.22}{266/500}12.8\% +
234\frac{0.78}{234/500}8.1\%}\biggr)\\
&&{}\Big/\biggl({266\frac{0.22}{266/500} +
234\frac{0.78}{234/500}}\biggr) = 9.1\%.
\end{eqnarray*}

Similarly, $\hat{\alpha}_{1} = 4.8\%; \hat{\alpha}_{1}^{*} = 3.1\%$. For
the nonparametric approach presented in Section \ref{sec3.2}, the same
results are
obtained. Indeed, whether using the risk difference or the log-odds ratio
as metrics, the parametric approach presented in Section \ref{sec3.1}
and the
nonparametric approach presented in Section \ref{sec3.2} lead to the
same results
for this example.

The standard error of the calibrated effect size $\hat{\alpha}_{j}^{*}$ can
be calculated directly here as
\begin{eqnarray*}
\operatorname{std}\bigl(\hat{\alpha}_{0}^{*}\bigr) &=& \sqrt{0.22^{2}
\times\frac{12.8\% \times
( 1 - 12.8\% )}{266} + 0.78^{2} \times\frac{8.1\% \times (
1 - 8.1\% )}{234}} \\
&=&
0.015.
\end{eqnarray*}

Similarly, $\operatorname{std}(\hat{\alpha}_{1}^{*}) = 0.005$. We could also use
statistical software to obtain the standard error. In this paper we used
the SAS procedure PROC GEMMOD with the generalized estimating equation
(GEE) option to compute the standard error described in Theorem \ref{theo1}. The
resulting standard errors for $\hat{\alpha}_{0}^{*}$ and
$\hat{\alpha}_{1}^{*}$ are 0.015 and 0.005, which are the same as obtained
in our direct computation.

Now, let us quantify the treatment effect using the log-odds ratio metric.
The estimate can be obtained through PROC NLMIXED with the replicate
statement. However, the standard error obtained from this procedure is not
the standard error stated in Theorem \ref{theo1}. In order to obtain the correct
standard errors, we could obtain a bootstrap standard error [\citet
{Efr81}],
which is 0.25. Alternatively, we can also use PROC GEMMOD to obtain the
same point estimate and the standard error. It results in the same estimate
and standard error. Note that the unadjusted log-odds ratio is
$\hat{\mu}_{\CP}=0.86$ with standard error 0.21.

The estimated log-odds ratio of Palivizumab vs. Motavizumab is 0.31
with a
standard error of 0.20. Using the unadjusted or adjusted effect size, we
calculate the unadjusted and adjusted statistics,
\begin{eqnarray*}
\hat{\mu} &=& \frac{\hat{\mu}_{\TC} + \hat{\mu}_{\CP}}{\sqrt{\sigma_{\TC
}^{2} +
\sigma_{\CP}^{2}}} = \frac{0.31 + 0.86}{\sqrt{0.20^{2} +0.21^{2}}} =
4.0;\\
\hat{\mu}_{\mathrm{adj}} &=& \frac{\hat{\mu}_{\TC} +
\hat{\mu}_{\CP}}{\sqrt{\sigma_{\TC}^{2} + \sigma_{\CP}^{2}}} = \frac
{0.31 +
1.14}{\sqrt{0.20^{2} +0.25^{2}}} = 4.5.
\end{eqnarray*}

The significance levels associated with the unadjusted and adjusted
inference are 0.00003 and 0.000003, respectively. Other than this approach,
one can also use a more conservative approach, the fixed margin
approach [see \citet{autokey7} for details], to make the following inference:
\begin{eqnarray*}
\hat{\mu}_{f} &=& \frac{\hat{\mu}_{\TC} + \hat{\mu}_{\CP}}{\sigma_{\TC} +
\sigma_{\CP}} = \frac{0.31 + 0.86}{0.20 +0.21} = 2.9;\\
\hat{\mu}_{\mathrm{adj},f} &=& \frac{\hat{\mu}_{\TC} + \hat{\mu}_{\CP}}{\sigma_{\TC} +
\sigma_{\CP}} =
\frac{0.31 + 1.14}{0.20 +0.25} = 3.2.
\end{eqnarray*}

The significance level associated with the unadjusted inference is 0.002
and it is 0.0006 for the adjusted inference. Although both are less than
0.05, the latter one is approximately $0.025^{2}$, which means the
significance level is as low as that of two independent clinical trials,
each significant at a level of 0.025, fulfilling the regulatory requirement
on the quantity of the evidence [see \citet{Sooetal13}]. The quantity
requirement has been interpreted in the FDA guidance of drug effectiveness
[\citet{autokey6}] as follows: ``With regard to quantity, it has
been FDAs
position that Congress generally intended to require at least two adequate
and well-controlled studies, each convincing on its own, to establish
effectiveness.'' Therefore, the adjusted approach could make a difference
because Motavizumab was evaluated in a single noninferiority trial.
However, we emphasize that this analysis only takes the published data into
consideration and assumes that there are no other potential issues
associated with the trial design and conduct.

\subsection{Calibrating the effect size using subject-level
data}\label{sec4.3}

Section \ref{sec4.2} presented a simple example to illustrate
calibration through
likelihood reweighting. In general, subject-level data from a clinical
trial is much more complex than the data presented in Table \ref{tab1}. Although the
FDA typically has access to all subject-level data for regulatory purposes,
we have no authority to use them for purposes other than regulatory
decision-making. Therefore, we cannot share our experiences analyzing real
data with readers. For the purpose of illustrating methodology, we will use
a simulated data set based on the IMPACT data set and the MOTA data set.

We randomly generate variables $\mathrm{x}_{1}$, $\mathrm{x}_{2}$, and
$\mathrm{x}_{3}$ so the generated data set, say, IMPACT$_{0}$ and
MOTA$_{0}$, will have 5 variables: BPD status, treatment, $\mathrm{x}_{1}$,
$\mathrm{x}_{2}$, and~$\mathrm{x}_{3}$. In IMPACT$_{0}$, we randomly
generate a data set for 1502 subjects, with $\mathrm{x}_{1}$,
$\mathrm{x}_{2,} \mathrm{x}_{3}$ following three independent Bernoulli
distributions with success rates of 0.4, 0.6, and 0.5. Similarly, in
MOTA$_{0}$, we randomly generate another data set for 6635 subjects, with
$\mathrm{x}_{1}$, $\mathrm{x}_{2}$, and $\mathrm{x}_{3}$ following three
independent Bernoulli distributions with success rates of 0.6, 0.5, and
0.4. We then pool the two data sets together and define a trial indicator
to distinguish IMPACT$_{0}$ and MOTA$_{0}$.

To obtain the weight for each subject, we use logistic regression to model
the logit of the trial probability as a linear function of BPD status,
$\mathrm{x}_{1}$, $\mathrm{x}_{2,}$ and $\mathrm{x}_{3}$. Using the fitted
model, we predict the probability of each subject in IMPACT$_{0}$ being
located in MOTA$_{0}$. We then define the weight $r(x)$ by the odds of the
predicted probability times $1502/6635$. Note here vector x means all
variables: BPD status, treatment, $\mathrm{x}_{1}$, $\mathrm{x}_{2,}$ and
$\mathrm{x}_{3}$. Now, the estimated propensity score ratio $r(x)$ is
defined for all 1502 subjects.

The MLE of the reweighted likelihood (\ref{equZV}) can be obtained through
implementation of SAS procedure PROC GEMMOD (see the attached programming
code). With the repeat and weight statement [in the repeat statement,
subject is the ID number and the weight is $r(x)$], the GEMMOD procedure
provides the GEE [\citet{ZegLia86}] type sandwich estimates for
standard error, corresponding to the variance formula given in Theorem \ref{theo1}.
All the programs, including the simulated data, are available upon
request.

The point estimate of the adjusted log-odds ratio is 1.23 with a standard
error of 0.28. The final inference of the noninferiority trial may use the
following adjusted statistics:
\begin{eqnarray*}
\hat{\mu}_{\mathrm{adj}} &=& \frac{\hat{\mu}_{\TC} +
\hat{\mu}_{\CP}}{\sqrt{\sigma_{\TC}^{2} + \sigma_{\CP}^{2}}} = \frac
{0.31 +
1.23}{\sqrt{0.20^{2} +0.28^{2}}} = 4.5;\\
\hat{\mu}_{\mathrm{adj},f} &=& \frac{\hat{\mu}_{\TC} + \hat{\mu}_{\CP}}{\sigma_{\TC} +
\sigma_{\CP}} =
\frac{0.31 + 1.23}{0.20 +0.28} = 3.2.
\end{eqnarray*}

Although the problem did not occur in our example, we note that the
weighted likelihood approach may result in an estimate with large
variance if we have very small or very large values of $r(x)$. The
problem has been described in the propensity score literature and is
not unique to our setting. Some promising methods for dealing with
extreme propensity score weights include using generalized boosted
regression (GBR) [\citet{McCRidMor04}, \citet{RidMcC} and
\citet{LeeLesStu09}]. Based on our experience, we recommend that the
propensity score model should be limited to effect modifiers, that is,
baseline variables that are associated with the treatment difference,
echoing what was recommended by \citet{ColHer08}. Including many
variables that are not effect modifiers in the propensity score model
generally increases the chance of extreme weight due to the larger
population heterogeneity, with no clear benefits in reducing bias. The
supplement [\citet{NieZhaRub}] provides more details and a
simulation study to illustrate this recommendation. At the end of this
section, we also describe some additional alternatives approaches.
Before doing that, we would like to discuss some special issues when
this problem occurs in historically controlled trials and
noninferiority trials.\looseness=-1

When the weights are extremely small, such as when some subjects with
certain characteristics in the historical trial have no or few counterparts
(subjects with the same characteristics) in the noninferiority trials,
$r(x)$ is 0 or near 0 for these subjects. For example, this may happen when
more stringent inclusion criteria are implemented so that subjects with
less severe disease conditions at baseline were included in the historical
trials but are excluded from the noninferiority trial. It is understandable
these subjects (e.g., with less severe disease condition) may not
always be
used to make inferences about the control vs. placebo (i.e.,
$\hat{\mu}_{\CP}$) in the noninferiority trials subjects (e.g., with more
severe disease condition) unless we make an assumption that the treatment
effect $\hat{\mu}_{\CP}$ in these subjects dose not depend on baseline
disease status. Only with this assumption, we may multiply their weight
$r(x)$ using a large number so that these subjects still represent the
subjects with different characteristics. Without this assumption, the
method expectedly leads to a relatively larger variance because we discard
a portion of information from the historical trial.

The weights can be extremely large, such as when a subpopulation presented
in a noninferiority trial is not well represented in the historical trial.
For example, some subjects in noninferiority trials of HIV may use a newly
approved potent background drug that was rarely used or never used in the
historical trials. In this case, using historical data from another group
(subjects who did not have the new background drug) to make inference about
the relative effect of the control vs. placebo
(i.e., $\hat{\mu}_{\CP}$) may not be prudent without additional
assumptions. In this case, we might have to consider alternative
approaches. One possibility is to restrict the proposed analysis to the
subpopulation of the current study that is also represented in the
historical study. This is essentially equivalent to what propensity score
matching would achieve, where unmatched subjects are automatically
excluded. Another possibility is to consider the hybrid design idea
presented in \citet{Sooetal11}.

Now we briefly describe a stratified approach based on stratified
propensity scores. Suppose a control treatment is evaluated in historical
trials but we would like to calibrate its effect size in a new population
for which $r(x)$ is computed. We group $r(x)$ into a number of strata
$g_{1},\ldots,g_{L}$. The percentage of subjects falling into $g_{l}$ is
$w_{lh}$ and $w_{ln}$ in population $P$ and $P^*$, respectively. Let
$\hat{\beta}_{l}$ with variances $s_{l}^{2}$ be the treatment effect
size in
group $l$. Then a combined calibrated treatment effect in the population
$P^*$ is $\sum_{l = 1}^{L} \hat{\beta}_{l}\mathrm{w}_{\ln}$ with variance
$\sum_{l = 1}^{L} s_{l}^{2}w_{ln}^{2}$. When a new treatment is evaluated
by its comparison to the control in population $P^*$, a stratified analysis
based on the subclasses of the propensity score can be implemented as
follows. Let $\hat{\gamma}_{l}$ be an estimate of treatment effect of the
new treatment, with variances $s_{\ln}^{2}$, in the lth group of the
propensity score. We evaluate the new treatment through a stratified
analysis such as
\[
\frac{\sum_{l = 1}^{L} \{ \hat{\gamma}_{l} - \hat{\beta}_{l} \}
w_{l^*}}{\sum_{l = 1}^{L} \{ s_{\ln}^{2} + s_{lh}^{2} \}
w_{l^*}^{2}}.
\]

Depending on the objective, $w_{l^*}$ may be chosen differently.
Alternatively, we may also use the stratification method as described in
Section \ref{sec4.3} or define a threshold $\Delta_{2} > \Delta_{1} >
0$ so that
subjects with $r(x)$ beyond [$\Delta_{1},\Delta_{2}$] should redefine the
weight of [$\Delta_{1},\Delta_{2}$], whichever is closer, similar to the
method used in \citet{ColHer08}. The actual determination of
[$\Delta_{1},\Delta_{2}$] depends on the actual data and practical
assumptions.

\section{Concluding remarks}\label{sec5}

Motivated by a real example, we show that bias can arise in active
controlled noninferiority trials when estimated treatment effect size for
the control treatment is obtained from a historical trial that has been
conducted for a different population. Covariate adjustment approaches
[\citet{Zha09} and \citet{NieSoo10}] have been proposed to
address the
problem. However, they may be directly applied to obtain the marginal
treatment effects, which are often the pre-specified primary endpoints.
This paper proposes a likelihood reweighting method through propensity
scoring to estimate the marginal treatment effect size in the target
population of a noninferiority trial based on data obtained from a
historical trial that has been conducted for a different population.

%
\begin{appendix}\label{app}
\section*{Appendix}

\begin{pf*}{Proof of Theorem \ref{theo1}}
It is easy to verify that the generalized linear model satisfies
Assumptions A1--A3 in \citet{Whi82}, therefore, Theorem 2.2 is
applicable. The log likelihood of (\ref{equZV}) is
\[
\sum_{i = 1}^{n} \frac{f^*(x_{i})}{f(x_{i})}\log
l_{t}(y_{it},\alpha_{t}).
\]
According to Theorem 2.2 [\citet{Whi82}], the maximum likelihood estimate
$\hat{\alpha}_{t}$ converges to the parameter, say, $\alpha_{t0}$, which
maximizes
\[
\int E_{Y|X = x,T = t} \bigl\{ \log f(x) + \log l(y_{it}|x,t) - r(x)
\log l(y_{it},\alpha_{t}) \bigr\} \,dF(x),
\]
where $\log l(y_{it}|x,t)$ is the log-likelihood function obtained from
model (\ref{eq1}). Taking derivatives with respect to $\alpha_{t}$, we know
$\alpha_{t0}$ is the solution of the following equation:
\[
\int E_{Y|X = x,T = t} \bigl[ r(x) \bigl\{ y - b' (
\alpha_{t} ) \bigr\} \bigr] \,dF(x) = 0.
\]
Consequently, the estimating equation can be written as
\[
\int E_{Y|X = x,T = t} \bigl\{ y - b' ( \alpha_{t} )
\bigr\} \,dF^*(x) = 0.
\]
Noting $b' (\cdot) = g^{ - 1}(\cdot)$, we have $\alpha_{t0}=\mu_{iP^{*}}
= g [ E_{X \in P^{*}} \{ \mu_{t}(X) \} ]$. As the
assumptions A4--A6 are easily verifiable, the asymptotic properties of the
MLE of (\ref{equZV}) are immediately obtained from Theorem 3.2 in \citet{Whi82}.
\end{pf*}
\end{appendix}

\section*{Acknowledgments}

We would like to express our great appreciation to the three reviewers,
an Associate Editor, and Dr. Paddock for all of their comments and
suggestions, which substantially improved the quality of the paper.

\begin{supplement}
\stitle{Supplement to ``Likelihood reweighting methods to reduce
potential bias in
noninferiority trials which rely on historical data to make
inference''}
\slink[doi]{10.1214/13-AOAS655SUPP} 
\sdatatype{.pdf}
\sfilename{aoas655\_supp.pdf}
\sdescription{The supplement provides an assessment of the efficiency
loss for the weighted likelihood method and a comparison between the
likelihood reweighting method and related methods in historically
controlled trials.}
\end{supplement}

%

\printaddresses

\end{document}